\documentstyle[twoside,fleqn,espcrc2,epsfig]{article}

\newcommand{\AmS}{{\protect\the\textfont2
  A\kern-.1667em\lower.5ex\hbox{M}\kern-.125emS}}

\title{Helioseismology and solar neutrinos: an update}

\author{G. Fiorentini
        , B. Ricci and  
        F.L. Villante
        \address{Dipartimento di Fisica, Universit\'a di Ferrara and INFN-Sezione di Ferrara,\\
        Via Paradiso 12, I-44100 Ferrara, Italy}}
       
\begin{document}

\begin{abstract}
We review  recent advances concerning helioseismology,
solar models and solar neutrinos. Particularly we  address the
following points:
i) helioseismic tests of recent SSMs;
ii) predictions of the Beryllium 
neutrino flux based on helioseismology;
iii) helioseismic tests regarding the
screening of nuclear reactions
in the Sun.
\end{abstract}

\maketitle

\section{Introduction}

Helioseismology has added important data on the solar
structure which provide severe constraint and tests
of SSM calculations. For instance, helioseismology 
accurately determines the depth of the convective zone $R_b$
and  the photospheric helium abundance $Y_{ph}$.
With these additional constraints there are essentially
no free parameters  for SSM builders.

In this paper we review  recent advances concerning helioseismology,
solar models and solar neutrinos. Particularly we shall address the
following points:
i) helioseismic tests of recent SSMs;
ii) prediction of the Beryllium neutrino flux based on helioseismology;
iii) helioseismic tests concerning the screening of nuclear reactions
in the Sun.

\section{A summary of helioseismic determinations of solar properties}

While we refer to e.g. \cite{eliosnoi} for a review of the
method and to \cite{referenze} for the data, we recall that by 
measurements of thousands of solar frequencies (p-modes) with a 
typical accuracy of $\Delta \nu/\nu \simeq 10^{-4}$, one derives:

\begin{figure}[htb]
\vspace{9pt}
\epsfig{file=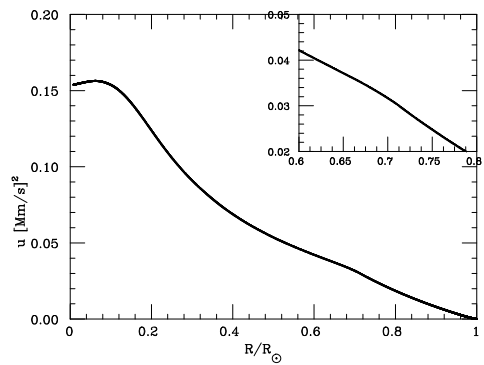,width=5.5cm,angle=90}
\caption{The isothermal sound speed profile, $u=P/\rho$,
as derived from helioseismic observations, 
from \cite{eliosnoi}.}
\label{fig1}
\end{figure}

\begin{itemize}
\item 
 {\it properties of the present convective envelope},  
such as depth  and helium abundance:
\begin{eqnarray}
\label{eqprop1}
R_b&=& 0.711 (1\pm 0.4\%) R_\odot  \\ 
\label{eqprop2}
Y_{ph}&=& 0.249 (1\pm 4\%)
\end{eqnarray}
The quoted errors, mostly resulting from systematic uncertainties in
the inversion technique, have been estimated conservatively by
adding linearly all known individual uncertainties, see \cite{eliosnoi}.
If uncertainties  are added in quadrature, the global error is about
one third  of that indicated 
in eqs. (\ref{eqprop1},\ref{eqprop2}), 
see again \cite{eliosnoi}.
This latter procedure was also used by Bahcall et al. \cite{bahbasu} 
with similar results. 
This yields the so called  ``$1\sigma$'' errors. We shall refer to the
conservative estimate as the ``$3\sigma$'' determination.
We remark however that this terminology is part of a slang,
and it does not correspond to well defined confidence levels,
as one has to combine several essentially systematic errors.
\item
{\it sound speed profile.}
 By inversion of helioseismic data  one can determine
the sound speed  in the solar interior. This analysis can be performed
in terms of either the isothermal sound speed squared, $u=P/\rho$, or in terms
of the adiabatic sound speed squared  $c^2= \partial P/ \partial \rho|_{ad}=
\gamma P/\rho$, as the coefficient
$\gamma= \partial \log P/ \partial \log \rho|_{adiab}$
is extremely well determined by means
of the equation of state of the stellar plasma.

In fig.~\ref{fig1} we show
the helioseismic value of $u$ as a function of the radial
coordinate $R/R_\odot$. The typical ``$3\sigma$'' errors
are of order $\pm 0.4\%$ in the intermediate solar region,
$R/R_\odot \simeq 0.4$, and increase up to $\pm 2\%$ near the
solar center.
\end{itemize}

\section{Helioseismic tests of  recent Standard Solar Models}

Fig.~\ref{fig2} compares the results of six different observational
determinations of the sound speed  with the results of the
best solar model of ref. \cite{bp00}, hereafter BP2000.
This figure suggests several comments:

\begin{figure}[htb]
\vspace{9pt}
\epsfig{file=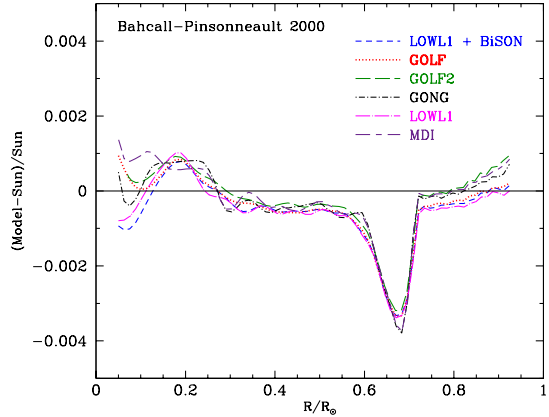,width=5.5cm,angle=-90}
\caption{Fractional difference between the sound speeds
calculated for the BP2000 model and the sound speeds in
six helioseismological experiments \cite{referenze}, from
\cite{bp00}.}
\label{fig2}
\end{figure}

i)Different measurements yield quite consistent value of the sound speed,
to the level 0.1\%;

ii) The solar model  BP2000 is in agreeement with helioseismic data 
to better than 0.5\% at any depth in the sun. We remark that
also the properties of the convective envelope predicted
by BP2000
($R_b/R_\odot=0.714, \, Y_{ph}=0.244$)
are in  agreement with helioseismic determinations, see eqs.
(\ref{eqprop1},\ref{eqprop2}).

iii) On the other hand, the predicted sound speed differs
from the helioseismic determination at the level of 0.3-0.4\%
{\it just below the convective envelope}.

\begin{figure}[htb]
\vspace{9pt}
\epsfig{file=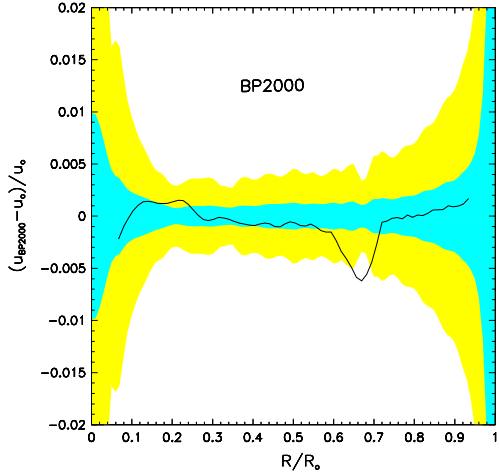,width=6.5cm,angle=0}
\caption{Comparison between  BP2000  
\cite{bp00} and MDI helioseismic data 
for $u=P/\rho$. The ``$1\sigma$'' and ``$3\sigma$'' 
helioseismic uncertainties \cite{eliosnoi} correspond 
to the dark and light areas respectively.}
\label{fig3}
\end{figure}

Concerning this last point, we remark that the difference is however
at the level of the ``$3\sigma$'' uncertainty of the 
helioseismic determination, as it is shown in
Fig.~\ref{fig3}. 
Nevertheless it can be taken as an indication of some imperfection of the
SSM. A marginally better agreement can be obtained in solar models including
mixing induced by rotation, see e.g. \cite{bp00} and in models including
macroscopic transport term, see e.g. \cite{tcz}.

In summary  all this means that SSM predictions are  accurate to the level
of one per cent or better, although there are indications of
some deficiencies at the level of per mille.

Concernig neutrino physics, the properties of the central
solar region, where nuclear reactions are efficient, are
relevant. Specifically, Boron and Beryllium neutrinos are produced very
near to the solar center with  maximal
production rates respectively at $R_{B}=0.04 R_\odot$ and
$R_{Be}=0.06 R_\odot$. Since the p-modes which are observed do not propagate
(actually are exponentially damped) so deeply in the sun the question
often arises if present helioseismic data can determine
the sound speed in region of Beryllium and Boron production.
From an extensive analysis of the inversion method and of data
available at that time we already concluded in \cite{eliosnoi}
that $u(R\simeq0)$ is determined with a ``$1\sigma$'' accuracy of 
1\%. This point was further elucidated in \cite{valencia99} where
a simplified analysis was presented in order to produce
convincing evidence that helioseismology fixes the sound
speed near the solar center with such an accuracy.

\section{Helioseismology and the beryllium neutrino flux}

As well known, the production of neutrinos from $^7{\rm Be}+e^{-}\rightarrow
^7{\rm Li}+\nu_{e}$  is an important item
in the context of the so called ``solar neutrino puzzle'' for
several reasons:\\
i) The result of Gallium experiments would be (partially) consistent
with the hypothesis of standard neutrinos  if  Beryllium neutrino
($\nu_{Be}$) production rate, $L(\nu_{Be})$, 
is suppressed by an order of magnitude with respect 
to the prediction of the Standard Solar Model (SSM), see e.g. 
\cite{where,report}.\\
ii) If one accepts neutrino oscillations as the solution of the
solar neutrino puzzle, the  determination of the neutrino mass matrix 
depends however on the predicted value of $L(\nu_{Be})$.\\
iii) Direct experiments aiming to the determination of the
 $\nu_{Be}$ signal are in preparation and of course
the interpretation of their result will rely 
on  $L(\nu_{Be})$ \cite{borexino,lens}.

The SSM prediction for Beryllium neutrinos \cite{bp00},
\begin{equation}
\label{eqlbessm} 
L(\nu_{Be})^{SSM} =1.3 \cdot 10^{37} \, (1\pm 9\%) 
{\mbox{ s$^{-1}$  ($1\sigma$)} }
\end{equation}
is very robust, much more than that of Boron neutrinos. However
any additional information which does not rely on SSM are
clearly welcome.

In ref. \cite{berillio} a lower limit  on the
Beryllium neutrino flux on earth was found,
$\Phi(Be)_{min} = 1\cdot 10^9$ cm$^{-2}$ s$^{-1}$,
in the absence of oscillations,
by using helioseismic data, the B-neutrino flux measured by
Superkamiokande  and the hydrogen abundance at the solar center $X_c$
predicted by Standard Solar Model (SSM) calculations.
We remark that this abundance is the only result of SSMs
needed for getting $\Phi(Be)_{min}$.
Lower bounds for the Gallium
signal, $G_{min}=(91 \pm 3) $ SNU, and for the Chlorine
signal, $C_{min}=(3.24\pm 0.14)$ SNU, 
in the absence of oscillations have also been derived.
They are about $3\sigma$ above the
corresponding experimental values,
$G_{exp}= (75\pm 5)$ SNU \cite{gallex,sage}
and $C_{exp}= (2.56\pm 0.22) $ SNU \cite{homestake}.
We remark that predictions for $X_c$ are very stable among
different (standard and non standard) solar models, see \cite{berillio}.
In fact $X_c$ is essentially an indicator of how much hydrogen
has been burnt so far. The stability of $X_c$ corresponds to
the fact that any solar model has to account for the same
present and time integrated solar luminosity.

In ref.~\cite{RV99} a step forward was made; a determination
of $L(\nu_{Be})$ was directly obtained by means of helioseismology 
without using additional assumptions.
The basic idea is the following. In the SSM the pp-II termination 
(which is the Beryllium producting branch)
accounts for an appreciable fraction of the 
$^4$He produced near the solar center.
If Beryllium  production is suppressed 
-- now and in the past -- less $^4$He would have been produced near
the center. As a consequence,
the molecular weight should decrease and  
the sound speed should increase in this region.
In other words, we know that SSM calculations are in 
good agreement with helioseismology and we can expect 
that this agreement is spoiled if $\nu_{Be}$ production 
is substantially altered.

\begin{figure}[htb]
\vspace{9pt}
\epsfig{file=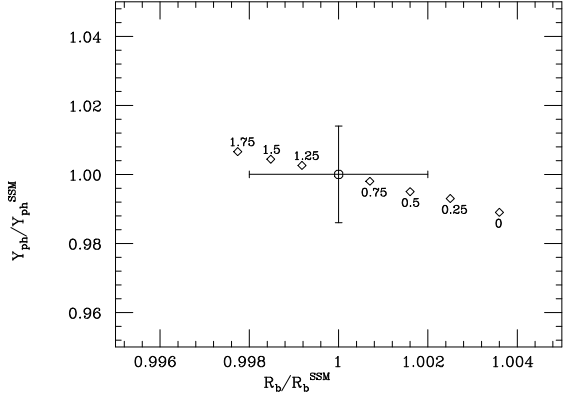,width=5.2cm,angle=90}
\caption{The photospheric helium abundance $Y_{ph}$ and  
the depth of the convective envelope $R_b$ 
in solar models with the indicated
values of $s=S_{34}/S_{34}^{SSM}$.
The error bars  correspond to the 
``$1\sigma$'' helioseismic uncertainties, from ref.\cite{eliosnoi}.}
\label{fig4}
\end{figure}

In order to test this idea, solar models with artifically changed 
$\nu_{Be}$ production  were constructed.
An efficient way for producing arbitrary
variations of $L(\nu_{Be})$ is to vary 
the zero energy  astrophysical S-factor ($S_{34}$)
for the reaction $^3{\rm He}+^4{\rm He}\rightarrow ^7{\rm Be}+ \gamma$,
since, as well known (see e.g. \cite{CDF,report,libro}), 
the production rate $L(\nu_{Be})$ 
is directly proportional to $S_{34}$:
\begin{equation}
L(Be)/L(Be)^{SSM}\simeq S_{34}/S_{34}^{SSM} ~.
\end{equation}

We remind that $S_{34}$ is measured with an accuracy of about 
ten per cent, ($S_{34}^{SSM}=0.54 \pm 0.05$ KeVb \cite{exps34}). 
It has been  varied, however, well beyond its
experimental uncertainty in order to simulate 
several effects which have
been claimed to suppress $\nu_{Be}$ production, 
e.g. hypothetical plasma effects which could alter 
nuclear reaction rates. 
\begin{figure}[htb]
\vspace{9pt}
\epsfig{file=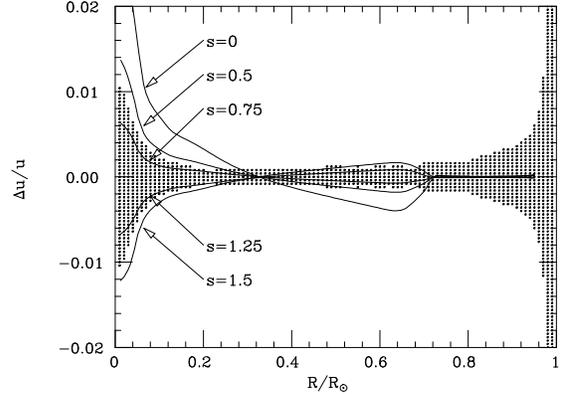,width=5.2cm,angle=90}
\caption{Fractional difference with respect to the SSM
prediction, (model-SSM)/SSM, of the isothermal squared sound
speed, $u=P/\rho$, in solar models with the indicated  
values of $s=S_{34}/S_{34}^{SSM}$. The dotted area corresponds
to the ``$1\sigma$''
 helioseismic uncertainty on $u$, from ref.\cite{eliosnoi}.}
\label{fig5}
\end{figure}
The resulting values for the quantities which can be tested 
by means of helioseismology are shown in Fig.~\ref{fig4}
and Fig.~\ref{fig5}, for several assumed values of the parameter
$s=S_{34}/S_{34}^{SSM}$.

The photospheric helium abundance $Y_{ph}$ is weekly sensitive
to the value of $S_{34}$ whereas the depth of the convective
envelope $R_b$ is altered 
by more than $1\sigma$ if $S_{34}$ is reduced below
one half of the SSM value, i.e.: 
\begin{equation}
L(\nu_{Be})=1.3 \cdot 10^{37} (1\pm 50\%) \quad
{\mbox{ s$^{-1}$ }} \,
{\mbox{at $1\sigma$}} ~. 
\end{equation}
As previously mentioned, one expects that the sound speed is altered, 
particularly near the solar center. In fact, stringent constraints 
arise from the sound speed profile,
particularly near $R\simeq 0.2 R_\odot$.
The requirement that the sound speed is not changed 
by more than $1\sigma$ yields:
\begin{equation}
L(\nu_{Be})=1.3 \cdot 10^{37} (1\pm 0.25)\quad
{\mbox{ s$^{-1}$ } }\,
 {\mbox{at $1\sigma$}}~. 
\end{equation}

In conclusion, 
helioseismology directly confirms
the production rate of  Beryllium neutrinos as
predicted by SSMs to within  $\pm 25\%$ ($1\sigma$ error).
This constraint is somehow weaker than 
that  estimated from uncertainties of the SSM,
see Eq.\ref{eqlbessm}, however it relies on direct observational 
data.

\section{Helioseismology and screening of nuclear reactions in the sun}

The study of screened nuclear reaction rates was started with the 
pioneering work of Salpeter 
\cite{salp}, who discussed both the extreme
cases of "weak" and "strong" screening,
providing suitable expressions for the screening factors
\begin{equation}
\label{eqf}f_{ij}= \langle\sigma v\rangle _{ij,plasma} 
/ \langle \sigma v \rangle _{ij,bare} \, .
\end{equation}
The solar core is  not far from the weak screening case,
however it does not satisfy the usual conditions under 
which the weak screening approximation holds.
This is the reason why the possibilities of large deviations
from weak screening have been investigated 
by several authors, 
expecially as an attemp to avoid or mitigate the 
``solar neutrino problem'', see e.g. \cite{b2000a} and references therein.

We remind that solar models are built by using stellar evolutionary 
codes which include specific expressions for the nuclear reaction rates.
If one uses  different formulas for the screening factors $f_{ij}$
one obtains different solar models.
On the other hand, helioseismology provides precise information
on the  sound speed profile and on the
properties of the convective envelope
which have to be reproduced by the correct solar model.
Therefore helioseismology can potentially provide a test of  
screening models.

In ref.~\cite{schermo}, solar models using different screening 
assumptions were constructed and compared with helioseismic data.
Specifically, four different model were considered:

\noindent  
i) The weak screening approximation (WES). 
The screening  factors $f_{ij}$  are given by:
\begin{equation}
\label{eqfweak}
\ln f_{ij}^{\rm{WES}} = Z_i Z_j e^{2} / (a_D\,kT)
\end{equation}
where $Z_i,Z_j$ are the charges of the interacting nuclei, 
$T$ is the temperature and $a_D$ is the Debye radius.
As clear from equation above,
 the screening factors are always larger than unity,
 i.e. the plasma provides enhancement of 
the thermonuclear reaction rates.

\noindent 
ii) The Mitler result \cite{mitler} (MIT), obtained with an analytical
method  which goes beyond the linearized approach and 
which correctly reproduces both the limits of weak and 
strong screening. 

\noindent
iii) Neglect completely any screening effect (NOS),
 i.e. nuclear reactions 
occur with rates $\langle\sigma v\rangle_{\rm{bare}}$.
This case is considered in connection with  the suggestions 
that screening can be much smaller than Salpeter's estimate,
see e.g. \cite{s95}.

\noindent 
iv) The Tsytovich model (TSY) \cite{tsy2000,bornatici}, which
provides a decrease of all the thermonuclear reaction rates with 
respect to the case of bare nuclei.

\begin{table}
\caption{Screening factors in  solar center, 
for weak screening (WES) \cite{salp}, 
Mitler model (MIT) \cite{mitler}, 
no screening (NOS) and Tsytovitch model
(TSY) \cite{tsy2000}.}
\begin{center}
\begin{tabular}{lcccc}
\hline
\hline
            & WES & MIT & NOS & TSY \\
\hline
$p+p$       &   1.049 &  1.045  & 1 & 0.949 \\
$^3He+^3He$ &   1.213 &  1.176  & 1 & 0.814 \\
$^3He+^4He$ &   1.213 &  1.176  & 1 & 0.810 \\
$^7Be+p$    &   1.213 &  1.171  & 1 & 0.542 \\
\hline
\hline
\end{tabular}
\end{center}
\label{tabf}
\end{table}

In Table \ref{tabf} the screening factors
at the solar center for the various models are shown.
One sees that  the weak screening approximation
always yields the largest enhancement factors,
as physically clear due to the fact that electrons 
and ions are assumed
to be free and capable of following
the reacting  nuclei.
By definition there is no enhancement in the NOS model, 
whereas in TSY model there is a decrease
of the reaction rate, as already remarked.  

\begin{figure}[htb]
\vspace{9pt}
\epsfig{file=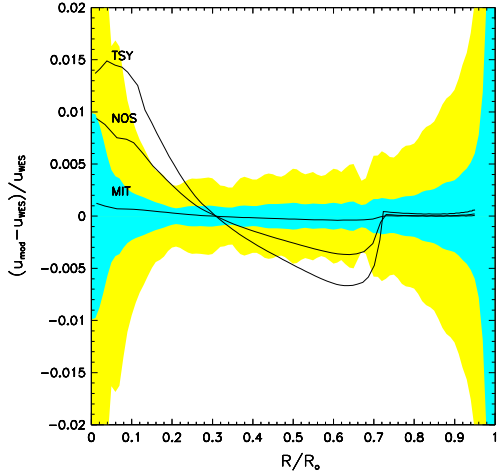,width=6.5cm,angle=0}
\caption{Comparison of different screening models
with the WES model for  $u=P/\rho$.
The ``$1\sigma$'' and ``$3\sigma$'' helioseismic uncertainties
\cite{eliosnoi} correspond to the dark and light areas.}
\label{fig6}
\end{figure}

Recent Standard Solar Models calculated by using the weak screening
prescription are in agreement with helioseismic constraints
on the properties of the convective envelope and on the sound speed
profile, see fig.~\ref{fig3}
This shows that the weak screening model is in agreeement with data
and deviations from WES cannot be too large.
From the comparison of different models, see 
Table \ref{tabmod} and fig.~\ref{fig6}, one obtains  
the following results: 


\begin{table}
\caption{Comparison among solar  models with different screening 
factors. We show the  fractional differences, (model -WES )/WES,
for the photospheric helium abundance ($Y_{ph}$), depth of the convective
envelope ($R_b$), central temperature ($T_c$), isothermal
sound speed squared at the solar center ($u_c$),
 neutrino fluxes ($\Phi_i$)
and predicted signals for Chlorine (Cl) and the Gallium (Ga) experiments.
\underline {All variations are in per cent}.}
\begin{center}
\begin{tabular}{lr@{}lr@{}lr@{}lr@{}}
\hline
\hline
&\multicolumn{2}{c}{MIT}
  &\multicolumn{2}{c}{NOS}
     &\multicolumn{2}{l}{TSY} \\
\hline
$Y_{ph}$                           &    -0.&076 
   &        -0.& 86               &   -1. & 4 \\
$R_b$                               &  + 0.& 037
   &        +0. &34              &   +0.& 59 \\
$T_c$                               &  + 0.&45
   &       +0. & 54               & +1.& 4 \\
$u_c$                               &  + 0.&10
   &       +1. & 0               & +1.& 4 \\
\hline
$\Phi_{pp}$                                &  +0.& 033
   &       +0. &45               & -0. &35  \\
$\Phi_{^7Be}$                              &   -0. & 19
   &        -2.  &4          &   -5.  &9 \\
$\Phi_{^8B}$                             &   -2. & 7
   &        -12. &                  &  +11 &   \\
\hline
Cl                                  &   -2. & 5
   &         -11. &                  & +9.&7 \\
Ga                                  & -0. &76
   &       -2. &9               & +2. &3 \\
\hline
\hline
\end{tabular}
\end{center}
\label{tabmod}
\end{table}

i) The difference between the Tsytovitch model (TSY) and 
the weak screening model (WES) exceeds the ``conservative" 
uncertainty on $u$ in a significant portion of the solar profile.
We remark that also the depth of the convective envelope is
significantly altered. In other words the anti-screening predictions
of ref. \cite{tsy2000,bornatici}
can be excluded by means of helioseismology.

ii) Also the difference between the no-screening model (NOS) and WES 
is significant for both $u$ and $R_b$ in comparison with 
helioseismic uncertainty. In other words the existence of 
a screening effect can be proved by means of helioseismology. 

iii) The Mitler model of screening (MIT) cannot be distinguished
from the weak screening model within the present accuracy of 
helioseismology.

\section{Concluding remarks}

We summarize here that main points of our discussion:

\noindent i) The comparison of recent SSMs with helioseismic
measurements shows that SSMs prediction are accurate to the 
level of one per cent or better, although there are
indications of some deficiencies at the level of
per mille.

\noindent ii) Helioseismology directly confirms
the production rate of  Beryllium neutrinos as
predicted by SSMs to within  $\pm 25\%$ ($1\sigma$ error). 

\noindent iii) Models for screening of nuclear reactions
in the Sun can be tested by means of helioseismology.
In particular, the anti-screening predictions of 
\cite{tsy2000,bornatici} can be excluded. In addition,
the existence of a screening effect can be proved by
helioseismology.


\end{document}